\documentclass[preprint,amsmath,amssymb, aps, pre]{revtex4-1}
\usepackage[english]{babel}
\usepackage[cp1251]{inputenc}
\usepackage{xcolor}
\usepackage{empheq}
\usepackage{graphics}
\usepackage{indentfirst} %
\usepackage[format=plain, justification=raggedright, font=small]{caption}
\usepackage{subcaption}

\everymath{\displaystyle}

\newcommand{\ci}{\mathrm{i}}
\newcommand{\phih}[1]{\hat{\varphi}_{#1}}

\graphicspath{{pictures/}}

\begin{document}

\title{Out-of-equilibrium one-dimensional disordered dipole chain}

\author{Anton V. Dolgikh}
\affiliation{Universite Libre de Bruxelles, Campus Plaine, CP231, Boulevard du Triomphe, 1050 Bruxelles, Belgium}
\affiliation{Mathematical Physics Department, Voronezh State University, Universitetskaya sq. 1, Voronezh 394006, Russia}

\author{Daniel  S. Kosov}
\affiliation{School of Engineering and Physical Sciences, James Cook University, Townsville, QLD, 4811, Australia}

\date{\today}

\begin{abstract}
We consider a chain of one-dimensional  dipole moments connected to two thermal baths with different temperatures. The system is in nonequilibrium 
steady state  and  heat flows through it. Assuming that  fluctuation of the dipole moment is a small parameter, we develop an analytically solvable model for the problem. The effect of disorder is introduced by randomizing the positions of the dipole moments. We show that the disorder leads to Anderson-like transition from conducting to a thermal insulating state of the chain. It is shown that considered chain supports both ballistic and diffusive heat transports depending on the strength of the disorder. We demonstrate that nonequilibrium leads to the  emergence of the long-range order between dipoles along the chain and make the conjecture that the interplay between nonequilibrium and next-to-nearest-neighbor interactions results in the emergence of long-range correlations in low-dimensional classical systems.
\end{abstract}
\maketitle

\section{Introduction}

The understanding of out-of-equilibrium  low-dimensional systems has been a challenging problem for decades. This topic covers a large variety 
of important problems of modern physics concerning, for example, the necessary conditions for the observation of the Fourier law~\cite{Bonetto2000, Dubi2009, Dubi2009B,Dubi2011}; how to achieve and manipulate directed transport in systems with Brownian motion~\cite{Reimann2002,Hanggi2009}; how to gain a  useful work in nonequilibrium~\cite{Reimann2002, Gelin2008}; how to control the energy transport in one- and  two-dimensional assemblies of large  organic molecules with high dipole moment arranged on a surface~\cite{Kottas2005,Ratner2007,Horansky2006}; the mechanism of the transition to  chaos in nonlinear chains~\cite{Ford1992}; the necessary conditions for the occurrence and existence of the temporally periodic and spatially  localized excitations in nonlinear chains~\cite{Aubry2006}; the unique steady state's existence in nonlinear chains~\cite{Eckmann1999}. Nonequilibrium  processes in low-dimensional systems are also of practical and technological interest because of the recent advances in 
nanofabrication.

In this paper, we consider an out-of-equilibrium one-dimensional chain of particles interacting with each other via the classical dipolar 
potential. Our interest is twofold. First, we consider a heat conduction in such chains. Second, we study the emergence of new correlations in the system caused by a heat flow.

We approximate a particle by  a point dipole placed at a fixed position on a line. This approximation is valid for different real physical systems. For instance,  molecules of artificial molecular rotors contain one or several chemical groups with substantial dipole moment. While the rest of the molecule is kept fixed on a surface these groups rotate~\cite{Kottas2005, Horansky2006}. The dipole chain  model is also intimately connected with ferrofluids or solid-state magnetic dipoles~\cite{Shelton2005, Murashov2002,Klapp2005}.  The point dipole approximation is valid for a 
single-file water chain~\cite{Hummer2001, Kofinger2008}. The water in  narrow single wall carbon nanotubes forms a strongly ordered one-dimensional chain~\cite{Hummer2001,Kyakuno2011}. Each water molecule is connected by two hydrogen bonds to neighbor molecules, instead of four bonds in bulk, and also interacts with the carbon atoms of the nanotube. The latter interaction is weak compared to the dipolar one, but owing to the high density of the carbon atoms, it is not negligible and leads to the additional stabilization of the water in  nanotube~\cite{Alexiadis2008,Kofinger2011}. The hydrogen bonds in 1D water are energetically stronger and possess longer life-time than ones in a bulk~\cite{Hummer2001}.  Both of these facts lead to the formation of the stable water molecules chain in narrow carbon nanotubes or pores~\cite{Hummer2001,Kofinger2009}. Moreover, equations of motion of dipole chain can be treated as a particular case of Kuramoto model~\cite{Kuramoto1984,Acebron2005}. In its turn, this model describes 
in a broad sense the synchronization in dynamical system of coupled oscillators. It is clear that this formulation is related to a number of phenomena from synchronization of cells in human organism to phase transitions or Brownian motors. The systems with dipole-dipole interaction occur not only in a classical physics, but also play an important role in quantum world. For instance, in a rapidly growing area of cold dipolar atomic gases~\cite{Lahaye2009,Arkhipov2005}. For example, in the high density limit single-file quantum dipoles can also form a lattice with a strong localization of atoms near lattice sites~\cite{Arkhipov2005}.

The remainder of the paper is organized as follows. In Sec.~\textrm{II}, we describe the Langevin nonequilibrium dynamics for the one-dimensional chain of 
dipoles, theoretical approach to treat the disordered and the method to compute nonequilibrium correlation functions. Section~\textrm{III} presents the results of the 
numerical calculations. Conclusions are given in Sec.~\textrm{IV}. Some technical aspects are relegated to appendices.

\section{\label{sec:Math formalism}Mathematical formalism}
\begin{figure}
\includegraphics[scale=1.0]{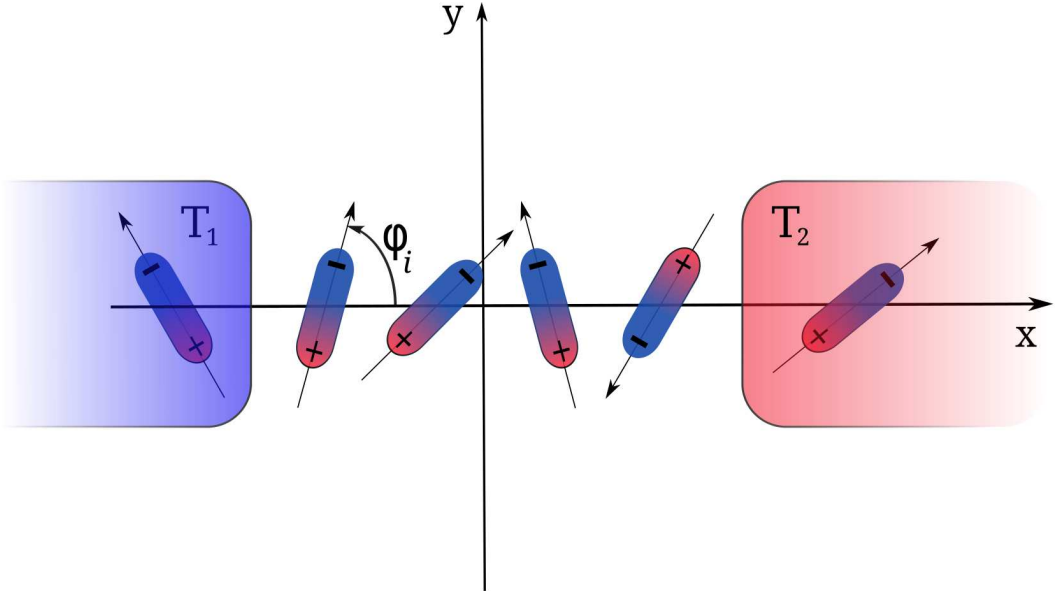}
\caption{\label{fig:dipole model} The sketch of the 1D dipole chain. Dipoles are represented by vectors with appropriate magnitude and orientation. In 
polar coordinates the state of the dipole is described by its magnitude and polar angle $\varphi$. There are two heat baths that support a heat flow 
along a chain. The left bath is the ''cold'' one, \textit{i.e.}, $T_L<T_R$.  }
 \end{figure}

\subsection{\label{subsec:Langevin dynamics} Nonequilibrium dynamics}
Let us consider a 1D chain of the $N$ point dipoles with magnitude $\mu$. The dipoles are rendered on a single line with a distance $a$ between them 
and are supposed to rotate in plane $(x,y)$ around the axes perpendicular to the plane of the dipole. The position of the arbitrary dipole is 
characterized by a single angle $\varphi$,~Fig.~\ref{fig:dipole model}. Thus, our model is purely one-dimensional.
The classical Hamiltonian of the dipole chain has the form:
\begin{eqnarray}
  H = K+U = \frac{I}{2}\sum\limits_{i=1}^N\dot{\varphi}_i^2 + \frac{\mu^2}{a^3}\sum\limits_{i=1}^N\sum\limits_{j=i+1}^{i+N_{cut}-1}\left(\frac{\sin{\varphi_i}\sin{\varphi_j-2\cos\varphi_i 
  \cos\varphi_j}}{|r_i-r_j|^3}\right), \label{eq:Hamiltonian}
\end{eqnarray}
where $I$ is the moment of inertia of the dipole, $\varphi_i$ is the angle between the vector of dipole moment of the $i$th dipole and the axis 
$x$, $r_i$ is the dimensionless position (in units of $a$) of the $i$th dipole moment and dot stands for time derivative, $N_{cut}=2,3,\ldots,N$ defines an interaction range and in what follows we call it cutoff radius.

The left and right dipoles of the chain are coupled by some mechanism to two macroscopic heat baths. The energy exchange between baths and system is 
implemented by Langevin dynamics~\cite{Lepri03, Dhar08}. The dimensionless (for units of measure see Appendix~\ref{sec:appA}) equations of motion have 
the following form:
\begin{equation}\label{eq:Langevin dynamic}
\begin{cases}
\dot{\varphi_{i}}=p_i, \\
\dot{p_i} = \displaystyle{-\frac12
\sum\limits_{j\neq i}^N\frac{3 \sin(\varphi_i+\varphi_j)+\sin(\varphi_i-\varphi_j)}{| r_i-r_j|^3}} + \delta_{i1}(\eta_L(t) - \gamma_L \dot{\varphi_i} 
) + \delta_{iN}(\eta_R(t) - \gamma_R \dot{\varphi_i} ),
\end{cases}
\end{equation}
where $\eta_L(t)$, $\eta_R(t))$ are the random forces of the left and right thermostats, respectively. We assume $\eta_{L,R}(t)$ to be the Gaussian 
white noise~\cite{Coffey2004}. Left and right viscosities $\gamma_L$, $\gamma_R$ are related to noises $\eta_L(t)$, $\eta_R(t)$ by the standard 
fluctuation-dissipation theorem:
\begin{equation}\label{eq:fluctuation dissipation}
 \langle \eta_{\{L,R\}}(t)  \eta_{\{L,R\}}(t')\rangle = 2\gamma_{\{L,R\}} T_{\{L,R\}} \delta(t-t').
\end{equation}

Let us assume that fluctuation of the dipoles near equilibrium positions are small enough to validate the expansion of sines in Eq.~\eqref{eq:Langevin 
dynamic} up to the term of the first order in $\varphi_i$. This condition is fulfilled in the case of strong dipolar interaction and moderate temperatures of the thermostats. In this approximation the 
equations of motion become:
\begin{equation}\label{eq:linearized Langevin dynamics}
 \begin{cases}
 \ddot{\varphi}_1 = -\varphi_1\sum\limits_{j=2}^N\frac{2}{|r_1-r_j|^3}-\sum\limits_{j=2}^N\frac{\varphi_j}{|r_1-r_j|^3}+\eta_L(t) - \gamma_L 
 \dot{\varphi}_1,\\
 \ddot{\varphi}_i = -\varphi_i\sideset{}{'}\sum\limits_{j=1}^N\frac{2}{|r_i- r_j|^3}-\sideset{}{'}\sum\limits_{j=1}^N\frac{\varphi_j}{|r_i- 
 r_j|^3},\\
 \ddot{\varphi}_N = -\varphi_1\sum\limits_{j=1}^{N-1}\frac{2}{|r_N-r_j|^3}-\sum\limits_{j=1}^{N-1}\frac{\varphi_j}{|r_N-r_j|^3}+\eta_R(t) - 
 \gamma_R\dot{\varphi}_N.
 \end{cases}
\end{equation}
Applying the Fourier transform to both sides of Eq.~\eqref{eq:linearized Langevin dynamics}, we arrive at the following linear system of equations:
\begin{equation}\label{eq:matrix equation for angles}
 \mathbf{M}\hat{\boldsymbol{\varphi}}=\boldsymbol{\eta},
\end{equation}
where $\hat{\boldsymbol{\varphi}} = \Big(\phih{1}, \ldots, \phih{N} \Big)^T$, $\boldsymbol{\eta}=\Big(\eta_L(\omega), 0,  \ldots, 0, 
\eta_R(\omega)\Big)^T$,  $\phih{i}=\int\limits_{-\infty}^{+\infty}\exp(\ci\omega t)\varphi_i(t)\, dt$,  and
\begin{equation}\label{eq:M-matrix}
 M_{ij}(\omega) = \begin{cases}
                   -\omega^2 + \sum\limits_{j=2}^N\frac{2}{|r_1-r_j|^3} - \ci \omega\gamma_L, & i=j=1, \\
                   -\omega^2 + \sideset{}{'}\sum\limits_{j=1}^N\frac{2}{|r_i-r_j|^3}, & i=j, \\
                   -\omega^2 + \sum\limits_{j=1}^{N-1}\frac{2}{|r_N-r_j|^3} - \ci \omega\gamma_R, & i=j=N,\\
                  \sideset{}{'}\sum\limits_{j=1}^N\frac{1}{|r_i- r_j|^3}, & i\neq j.
                  \end{cases}
\end{equation}
Now we are ready to calculate the steady-state energy current in system. To do this, let us consider the rate of energy change for the first dipole 
(connected to the left thermostat):
\begin{equation}\label{eq:energy rate of first dipole}
 \frac{d\epsilon_1}{dt} = \frac12p_1\sum\limits_{j=2}^N F_{1j} - \frac12\sum\limits_{j=2}^N F_{j1}p_j - p_1^2(t)\gamma_L + p_1 \eta_L(t),
\end{equation}
where $\epsilon_1$ is the energy density of the first dipole and $F_{1j}$ is the force that $j$th dipole exerts on first dipole. The last two terms in Eq.~\eqref{eq:energy rate of first dipole} are due to the thermostat. 
Thus, the energy current that flows into the system from the left thermostat is given by:
\begin{equation}
 j(t)=- p_1^2(t)\gamma_L + p_1 \eta_L(t) = -\dot{\varphi}_1(t)^2\gamma_L + \dot{\varphi}_1(t)\eta_L(t).
\end{equation}
The average current is:
\begin{equation}
 \langle j(t)\rangle=\langle -\dot{\varphi}_1(t)^2\gamma_L + \dot{\varphi}_1(t)\eta_L(t) \rangle.
\end{equation}
The function $\varphi_i(t)$ is found from Eq.~\eqref{eq:matrix equation for angles} by inverting the Fourier transform:
\begin{equation}\label{eq:angle via M-matrix}
\hat{\varphi_i}(\omega)=(M^{-1})_{i1}(\omega)\eta_L(\omega) + (M^{-1})_{iN}(\omega)\eta_R(\omega),
\end{equation}
and the expression for the steady state heat current becomes~\cite{Casher1971,Lepri03, Dhar2006, Dhar08}. 
\begin{equation}
\label{eq:current Langevin}
\langle j \rangle = \frac{2\gamma_L \gamma_R \Delta T }{2\pi} \int\limits_{-\infty}^{+\infty}\omega^2 |(M^{-1})_{1N}(\omega)|^2\,d\omega,
\end{equation}
where $\Delta T = T_R - T_L$. 

The integral in Eq.~\eqref{eq:current Langevin} is completely defined by the poles of the $(M^{-1})_{1N}$ element. By definition, poles of the 
$\mathbf{M}^{-1}$ are the roots of $\det \mathbf{M}(\omega)=0$ equation [details of the calculations of the integral and root finding of $\det \mathbf{M}(\omega)$ are given in  
Appendix~\ref{sec:appC}]. It follows from Eq.~\eqref{eq:M-matrix} that $\det \mathbf{M}(\omega)$ is the polynomial in $\omega$ of the order of $2N$. 
Therefore, according to the fundamental theorem of algebra, there are $2N$ solutions of this equation. These solutions correspond to the frequencies of 
elementary excitations that carry the energy along the chain. Solutions of  Eq.~\eqref{eq:linearized Langevin dynamics} for free dipole chain, not 
connected to the thermostats, have form $\varphi_j(t)=e^{\mathrm{i}( j k-\omega t)}$, where $k$ is the wave vector and $\omega$ is real. The 
interaction with thermostats leads to the occurrence of the imaginary part in $\omega$.

The dipoles are never strictly fixed at their positions under realistic conditions. Thermal fluctuations and other effects  are very difficult 
to eliminate completely. They lead to the fluctuations of the dipoles near their positions that brings the chain into a disordered state. To 
understand the role of the disorder we take into account the thermal fluctuations of dipoles in the direction along the line on which they are 
rendered. We adopt the simple scheme when dipoles' positions $r_i$ have Gaussian distribution
\begin{equation}\label{eq:Gaussian distribution for positions}
p(r_i) = \frac{1}{\sqrt{2\pi\sigma^2}}\exp\left(-\frac{(r_i-r_{i0})^2}{2\sigma^2}\right),
\end{equation}
where $r_{i0}$ is the position of the $i$th dipole in the ordered chain and $\sigma$ is the dispersion that characterizes the ''strength'' of the 
disorder. Heat current in a disordered chain is calculated from Eq.~\eqref{eq:current Langevin} by averaging over a number of realizations of the 
disorder.

\subsection{\label{subsec: corrfun calc} Correlation functions in nonequilibrium steady state}
The dipole orientation relaxation time is a measurable quantity in many experiments that allows to understand the physical nature of the processes 
under observation. The dipole-dipole correlation function is defined as $\langle \boldsymbol{\mu}(t)  \cdot  \boldsymbol{\mu}(0)  \rangle$, where 
$\boldsymbol{\mu}(t) = \sum\limits_{i=1}^N \boldsymbol{\mu}_i(t)$, $\boldsymbol{\mu}_i(t)$ is the dipole moment of the $i$th dipole and 
$\boldsymbol{\mu}(t)$ is total dipole moment of the chain (a polarization vector).
It follows from the definition of  the correlation function that:
\begin{equation}\label{eq:correlation function vs Delta}
\langle \boldsymbol{\mu}(0) \cdot \boldsymbol{\mu}(t) \rangle = \sum\limits_{i,j=1}^{N}\left( \Big \langle \cos \varphi_i(0)\cos \varphi_j(t) \Big\rangle + 
\Big\langle \sin \varphi_i(0)\sin \varphi_j(t) \Big\rangle \right)= \sum\limits_{i,j=1}^{N}\Big\langle \cos \Delta_{ij}(t) \Big\rangle,
\end{equation}
where  $\Delta_{ij}(t)=\varphi_j(t) - \varphi_i(0)$. It is known that if $X$ is the Gaussian random variable, then~\cite{Kampen1992}
\begin{equation}\label{eq:Gaussian variable identity}
\Big\langle \exp(\ci X) \Big\rangle = \exp\left(\ci\langle X\rangle -\frac12 \langle X^2 \rangle \right).
\end{equation}
One can see that according to Eq.~\eqref{eq:matrix equation for angles}, $\varphi_i$, $i=1,2,\dots, N$, are the Gaussian random variables with zero mean 
value $\langle \varphi_i\rangle=0$. Thus, we can use the rule Eq.~\eqref{eq:Gaussian variable identity} to calculate the average in Eq.~\eqref{eq:correlation 
function vs Delta}:
\begin{equation}\label{eq:correlation function pre calculate}
\begin{aligned}
\langle \boldsymbol{\mu}(0) \cdot \boldsymbol{\mu}(t) \rangle = & \sum\limits_{ij} \exp\left(-\frac12 \langle \Delta_{ij}^2(t)\rangle\right)=\\ 
=&\sum\limits_{ij}\exp\left(-\frac{\langle\varphi^2_j(t)\rangle+\langle\varphi^2_i(0)\rangle}{2}+\langle\varphi_j(t)\varphi_i(0)\rangle\right).
\end{aligned}
\end{equation}
It follows from Eq.~\eqref{eq:angle via M-matrix} that the first term in the exponent does not depend on time. The term $\langle\varphi_j(t)\varphi_i(0)\rangle$ is space cross-correlation function. It is evident from general considerations that
\begin{equation}
\langle\varphi_j(t)\varphi_i(0)\rangle  \xrightarrow[t\to+\infty]{}  \langle\varphi_j(t)\rangle\langle\varphi_i(0)\rangle=0.
\end{equation}
Hence
\begin{equation}
\langle \boldsymbol{\mu}(0) \cdot \boldsymbol{\mu}(t) \rangle \xrightarrow[t\to+\infty]{} 
\sum\limits_{ij}\exp\left(-\frac{\langle\varphi^2_j(t)\rangle+\langle\varphi^2_i(0)\rangle}{2}\right).
\end{equation}
For convenience, we subtract the right-hand-side of this limit from Eq.~\eqref{eq:correlation function pre calculate}. This will provide the dipole-dipole correlation 
function to vanish at $t\to+\infty$. Doing this, we get:
\begin{equation}
\langle \boldsymbol{\mu}(0) \cdot \boldsymbol{\mu}(t) 
\rangle=\sum\limits_{ij}\exp\left(-\frac{\langle\varphi^2_j(t)\rangle+\langle\varphi^2_i(0)\rangle}{2}\right)\Bigg[ 
\exp\left(\langle\varphi_j(t)\varphi_i(0)\rangle\right)-1\bigg] .
\end{equation}

The averages in this formula immediately follows from Eq.~\eqref{eq:angle via M-matrix}  :
\begin{equation}\label{eq:correlation function Langevin}
 \begin{aligned}
 \langle\varphi^2_n(t)\rangle = \langle\varphi^2_n(0)\rangle &= \frac{1}{(2\pi)^2}\int\limits_{-\infty}^{+\infty}\Big[ 2\pi\gamma_L T_L 
 M^{-1}_{n1}(\omega)M^{-1}_{n1}(-\omega) + 2\pi\gamma_R T_R M^{-1}_{nN}(\omega)M^{-1}_{nN}(-\omega)\Big]\,d\omega  \\
    \langle\varphi_n(t)\varphi_m(0)\rangle &= \frac{1}{(2\pi)^2}  \int\limits_{-\infty}^{+\infty}\exp(-\ci\omega t)  \Big[ 2\pi\gamma_L T_L M^{-1}_{m1}(\omega)M^{-1}_{n1}(-\omega) +\\
   \phantom{ \langle\varphi_n(t)\varphi_m(0)\rangle} &+ 2\pi\gamma_R T_R M^{-1}_{mN}(\omega)M^{-1}_{nN}(-\omega) \Big]\,d\omega.
  \end{aligned}
\end{equation}
The integrals are calculated by the same technique as the ones in Eq.~\eqref{eq:current Langevin}.

Finally, we can write the general form of the correlation function in the linearized Langevin dynamics:
\begin{equation}\label{eq:correlation function analytic}
\begin{aligned}
  &\langle \boldsymbol{\mu}(0) \cdot \boldsymbol{\mu}(t) \rangle =\sum\limits_{ij}^N K_{ij}\left[\exp\left( \sum\limits_{k=1}	^{2N} 
  G_{ij}(\omega_k)e^{-\mathrm{i}\omega_k t}  \right) - 1 \right]\\
  &K_{ij}= \exp\left(-\frac{\langle\varphi^2_j(t)\rangle + \langle\varphi^2_i(0)\rangle}{2}\right), \\
  &G_{ij}(\omega_k) = \frac{1}{2\pi}\left(\gamma_L T_L \frac{C_{i1}(\omega_k)C_{j1}(-\omega_k)}{\det'\mathbf{M}(\omega_k)\det\mathbf{M}(-\omega_k)} + \gamma_R 
  T_R \frac{C_{iN}(\omega_k)C_{jN}(-\omega_k)}{\det'\mathbf{M}(\omega_k)\det\mathbf{M}(-\omega_k)}\right), \\
  &\det\nolimits'\mathbf{M}(\omega_k)=\prod\limits_{n\neq k}(\omega_k - \omega_n),
\end{aligned}
\end{equation}
where $\omega_k$, $k=1,2,\ldots, 2N$ are zeros of $\det \mathbf{M}(\omega)$ lying in the lower half-plane of the complex plane $\omega$, $C_{ij}$ are the cofactors of the $\mathbf{M}(\omega)$. Coefficients $G_{ij}$ are calculated according the method described in Appendix~\ref{sec:appB}. We do not give the explicit form of $K_{ij}$ for brevity, but they can be represented in the same form as $G_{ij}$. It is seen that Eq.~\eqref{eq:correlation 
function analytic} decays exponentially only for times satisfying the condition $\mathrm{Im}\,\omega_k t \ll 1$.

The correlation function Eq.~\eqref{eq:correlation function analytic} can be considerably simplified if we expand the cosine in Eq.~\eqref{eq:correlation 
function vs Delta} into the power series in $\Delta_{ij}$ up to the second-order terms. In this case we have:
\begin{equation}\label{eq:correlation function linear analytic}
\begin{aligned}
&\langle \boldsymbol{\mu}(0)\boldsymbol{\mu}(t) \rangle \approx  \sum\limits_{ij} \langle \delta_{+}(t) - \frac12\Delta^2_{i,j}(t) 
\rangle=N^2\delta_{+}(t) -\frac12 \sum\limits_{ij}\langle \Delta^2_{ij}(t) \rangle,\\
&\langle \Delta^2_{ij}(t) \rangle = \langle\varphi^2_i(t)\rangle + \langle\varphi^2_j(0)\rangle -2\langle\varphi_i(t)\varphi_j(0)\rangle,
\end{aligned}
\end{equation}
where $\int\limits_{0}^{+\infty}\delta_{+}(t)=1$. The term $\langle \Delta_{ij}^2(t)\rangle$ is evaluated by a direct substitution 
from Eq.~\eqref{eq:correlation function Langevin}.  Thus, in this simple approximation the correlation function has exponential asymptotics,  $\langle 
\boldsymbol{\mu}(t)  \cdot  \boldsymbol{\mu}(0)  \rangle\sim\exp(-\ci \omega_k t)$ , where $\omega_k$ have the same meaning as above. Therefore, the 
longest relaxation time corresponds to the $\omega_k$ with the smallest absolute value of imaginary part lying in the lower half-plane $\omega$.

\section{Numerical results}
\subsection{\label{subsec:heat conduction numerical}Heat conductivity and temperature profile for disordered chain}
\begin{figure}
        \centering
        \begin{subfigure}[b]{0.5\textwidth}
                \includegraphics[width=\textwidth]{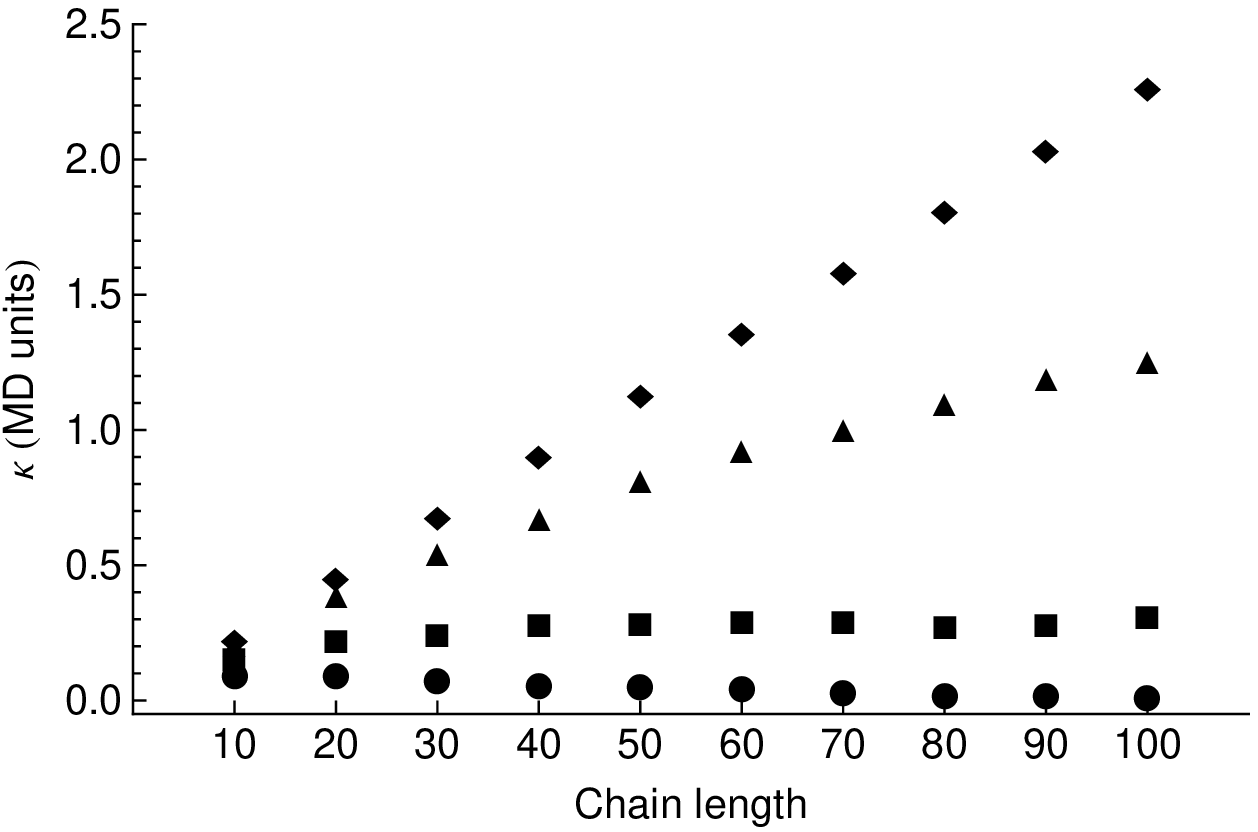}
        \end{subfigure}%
                 \begin{subfigure}[b]{0.5\textwidth}
                \includegraphics[width=\textwidth]{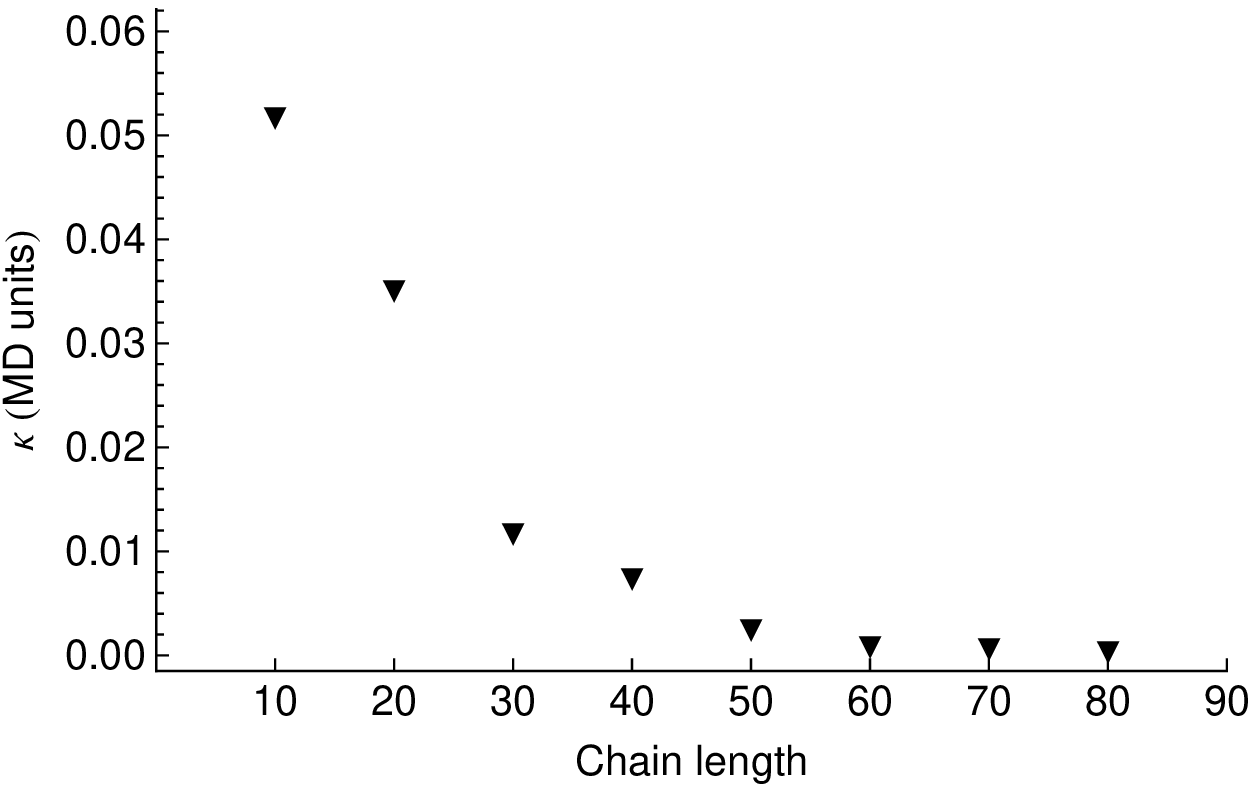}
       \end{subfigure}
        \caption{\label{fig:heat conductance disorder analytic} The heat conductivity of disordered dipole chain calculated according 
        to Eq.~\eqref{eq:current Langevin} for different disorder magnitudes, $T_L=0.19\,$MD units, $T_R=0.26\,$MD units, $N_{cut}=3$. On the left panel: $\blacklozenge$ -- $\sigma=0$, ordered chain; 
	$\blacktriangle$ -- $\sigma=0.02$; $\blacksquare$ -- $\sigma=0.05$; \textbullet$\;$ -- $\sigma=0.08$. Right panel presents the heat conductivity of disordered chain in case of strong dumping, $\sigma=0.11$. The results are obtained by averaging over $200$ realizations of disorder.}
\end{figure}

The quantitative measure of the heat transport in media is the heat conductivity $\kappa$.  In this article we adopt the ''global'' definition 
of the heat conductivity~\cite{Prosen2005} :
\begin{equation}\label{eq:heat conductivity global}
 \kappa(L) = \frac{j L}{\Delta T}, \; \Delta T = T_R - T_L,
\end{equation}
and $L$ is the chain length. Thus, to find $\kappa$ we first need to find heat current. The starting point of calculation of heat current 
is Eq.~\eqref{eq:current Langevin}. The calculations of heat current $j$ is straightforward for ordered dipole chain. In the case of disordered chain, we 
average a heat current over $200$ realizations of the dipole positions. The obtained dependence of the heat conductivity on the chain length is 
presented in Fig.~\ref{fig:heat conductance disorder analytic}.
It shows the transition from ballistic transport with infinite heat conductivity (for infinite chain) to diffusive transport with finite heat 
conductivity, and eventually the high level of disorder results into thermal insulating state. For ordered chain, $\sigma=0$, the heat conductivity is proportional to the length of the chain, $\kappa~\sim L$. For disordered ones 
$\kappa(L)$ law deviates from the linear dependence. The observed change of transport regimes is an example of a very general conductor-insulator 
transition induced by a disorder~\cite{Anderson1958, Resta2011}.

The question of interest is whether ballistic transport takes place as a result of linearization of the original equations of motion, Eq.~\eqref{eq:Langevin dynamic}, or if it is an intrinsic property of the model. To answer this question we  numerically integrated  nonlinear stochastic dynamical equation of motion Eq.~\eqref{eq:Langevin dynamic} by recently developed Langevin dynamics integrator~\cite{Bussi2007}. The time step in  numerical integrator is $0.03\,$MD units. First, we wait  for $10^8$ time steps to bring the system to the nonequilibrium steady state, then we perform the production run for additional $10^7$ time steps and computed the average heat flow.

\begin{figure}
\includegraphics[scale=0.7]{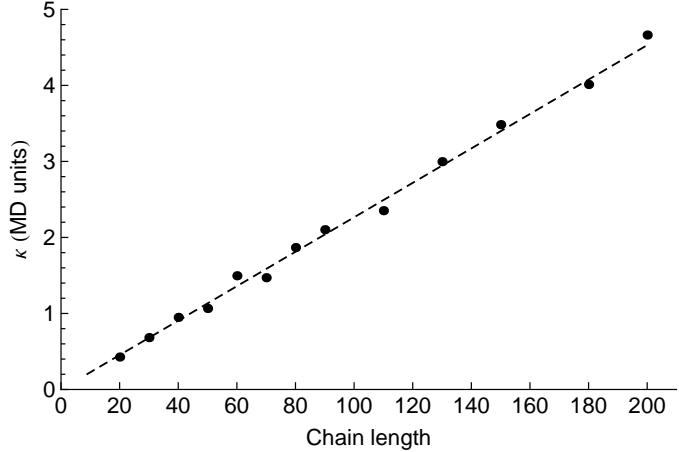}
 \caption{\label{fig:MD heat conductivity} Heat conductivity of the ordered dipole chain. Solid circles represent molecular dynamics simulation of the exact model Eq.~\eqref{eq:Hamiltonian}. The dashed line is $\kappa=j N/\Delta T$, where $j$ is calculated according to linearized approximation Eq.~\eqref{eq:current Langevin}.}
 \end{figure}

The results of the simulation are presented in Fig.~\ref{fig:MD heat conductivity}. We see that linearized model gives a very good approximation for the exact heat conductivity. This observation looks intriguing because it was stated recently~\cite{Dhar2008} that ''even a small amount of anharmonicity leads to a $j\sim 1/N$ dependence, implying diffusive transport of energy''. Nevertheless, Fig.~\ref{fig:MD heat conductivity} convincingly demonstrates the applicability of the linear approximation for the set of model's parameters used in our paper.

 \begin{figure}
\includegraphics[scale=0.9]{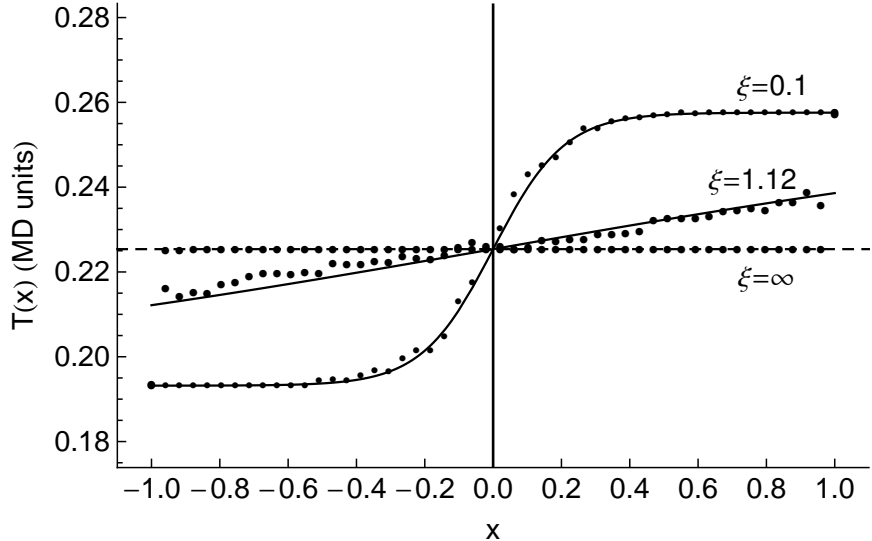}
 \caption{\label{fig:temperature profiles}  Temperature profiles for different disorder strengths and corresponding localization lengths according 
 to Eq.~\eqref{eq:temp profile approximation}. The disorders (in downward direction): $\sigma=0.18$, $\sigma=0.03$, $\sigma=0$.  Profiles in disordered 
 cases were obtained by averaging over $100$ disorder realizations. Dashed line shows position of the mean temperature $T(x)=(T_L+T_R)/2$. The meaning  of the localization length $\xi$ is revealed below in Eq.~\eqref{eq:temp profile approximation}.}
 \end{figure}

The disorder also affects the temperature profile of the chain. To illustrate this we calculated the temperature profile for different values of 
$\sigma$. We use the kinetic definition of local temperature, so the temperature $T_i$ of the $i$th dipole is given by
\begin{equation}\label{eq:temperature definition}
T_i=\langle \dot{\varphi}_i^2 \rangle.
\end{equation}
In Fig.~\ref{fig:temperature profiles} we show some temperature profiles corresponding to different disorder strength $\sigma$. For ordered chain the temperature profile coincides with the one in harmonic chain; \textit{i.e.}, the temperature of the internal dipoles is the average of the thermostats' temperatures, and the temperatures of the leftmost and rightmost dipoles are equal to the temperatures of the corresponding thermostats. Disorder destroys the flat 
temperature profile and leads to the formation of temperature gradient. The size of the region of the chain, where it occurs, depends on disorder strength. Such behavior of the temperature profile could be caused by the localization of the elementary excitations under the influence of disorder and in this respect it resembles the well known Anderson transition~\cite{Evers2008}.

The fundamental properties of heat conduction were considered recently, concerning with reconstruction of the Fourier's law in quantum 
wires~\cite{Dubi2009, Dubi2009B}. Under quite general physical assumptions a temperature profile was shown to have the form
\begin{equation}\label{eq:temp profile approximation}
 T(x)=T_L + \frac{\Delta T}{1+\exp(-x/\xi)},
\end{equation}
where $x=\frac{2n}{N-1}-1$, $n=0,2,\ldots, N-1$. Dubi and Di Ventra~\cite{Dubi2009} dubbed $\xi$ the ''thermal length'' because it characterizes the 
length-scale of the existence of a local temperature gradient. We approximate temperature profiles of our model by Eq.~\eqref{eq:temp profile 
approximation}. In Fig.~\ref{fig:temperature profiles} we see that increase of the disorder strength results in decrease of the $\xi$. According 
to~\eqref{eq:temp profile approximation} the case of the ordered chain corresponds to the infinite value of $\xi$. From Fig.~\ref{fig:temperature 
profiles} it is seen that temperature gradient in the system start to develop for values of $\xi$ of the order of a system size. For $\xi$ small enough compared to 
the system size one observes steep temperature gradient near the center of a chain while the parts of the chain close to edges are thermalized at the 
temperatures of the corresponding thermostat.

\subsection{Nonequilibrium correlation functions and relaxation time to the steady state}
\begin{figure}
\centering
\includegraphics[scale=0.8]{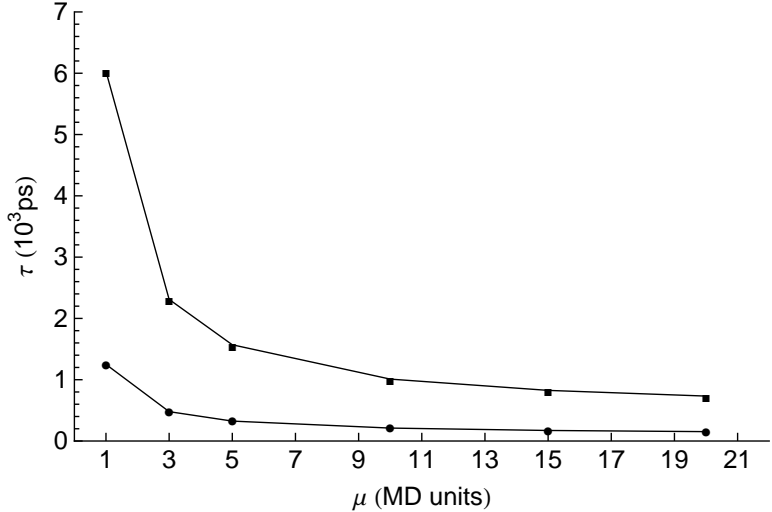}
\caption{\label{fig:relaxation times} Dipole relaxation times vs. dipole moment. The parameters are: $T_L=0.1\,$MD units, $T_R=0.3\,$MD units, $N_{cut}=3$; solid circles correspond to $N=30$, solid squares correspond to $N=50$.}
\end{figure}

Using Eq.~\eqref{eq:correlation function linear analytic}, we estimate the dipole relaxation times $\tau$ for different values of 
dipole moments $\mu$. The results presented in Fig.~\ref{fig:relaxation times} demonstrate inverse dependence $\tau(\mu)$. It has clear physical 
explanation. Namely, the time of the relaxation toward a nonequilibrium steady state is determined by the energy transfer inside a system and higher magnitude 
of dipole moments result in more effective energy transfer between dipoles and, consequently, relaxation to equilibrium state runs faster.

Two main factors influence the relaxation time in the dipole chain; see Fig.~\ref{fig:relaxation times}. The first one, is that the interaction strength between dipoles resulted in 
decreasing of relaxation times, and the second one, is that the chain length resulted in increasing of relaxation time.
To study the space correlations in chain we calculate the following correlation function $\langle \varphi_i(\infty)\varphi_j(\infty)\rangle$, where 
$\varphi_i(\infty)=\lim_{t\to+\infty}\varphi_i(t)$. Details on the calculation of space correlation function are given in Appendix~\ref{sec:appB}. Note, that instead of working with a number of correlation functions corresponding to different combinations of indexes $(i,j)$, we construct the function $C_n$:
\begin{equation}\label{eq:averaged space correlation function}
C_n = \frac{1}{N-n}\sum\limits_{i=1}^{N-n}\varphi_i \varphi_{i+n}.
\end{equation}
which expresses a ''smoothed'' behavior of space correlations.

\begin{figure}
        \centering
        \begin{subfigure}[b]{0.5\textwidth}
                \includegraphics[width=\textwidth]{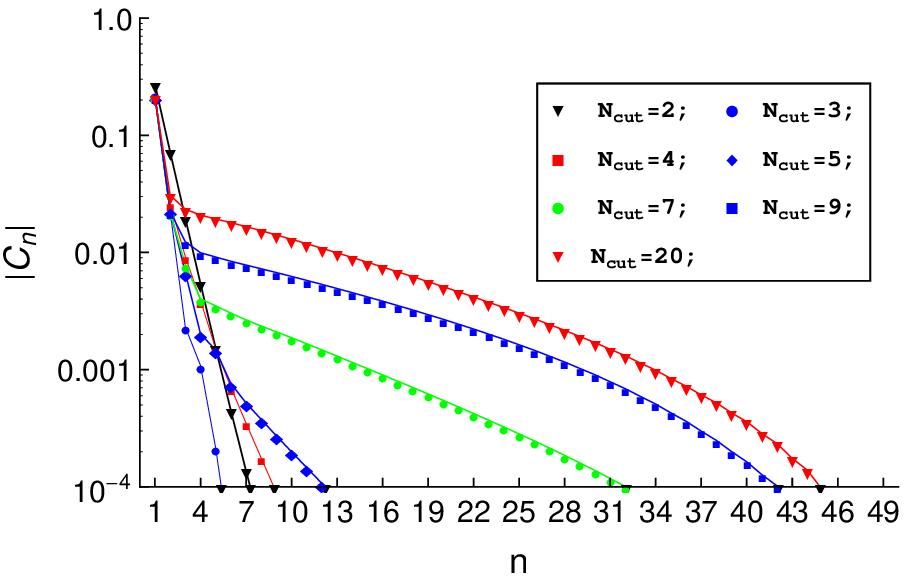}
        \end{subfigure}%
        \begin{subfigure}[b]{0.5\textwidth}
                \includegraphics[width=\textwidth]{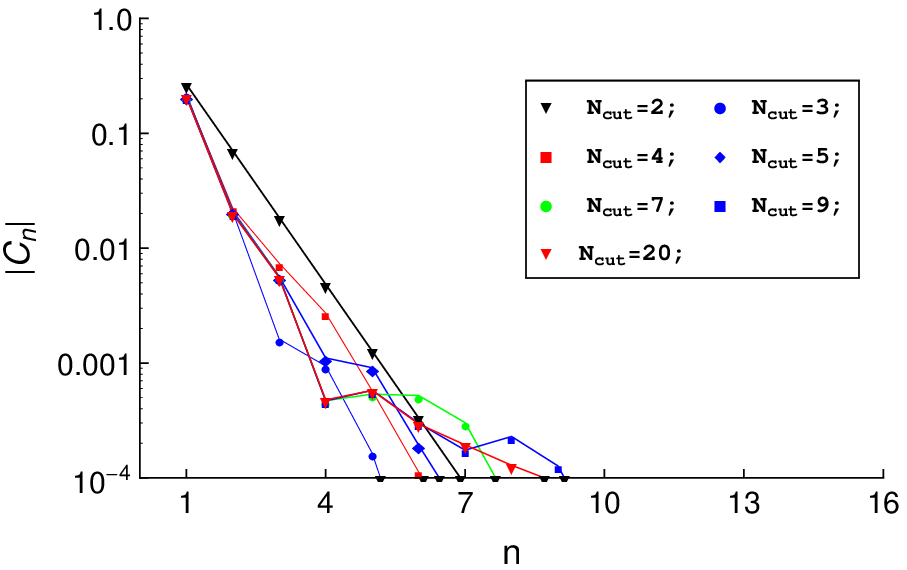}
       \end{subfigure}
\caption{\label{fig:space correalation functions comparison} The absolute values of the different space correlation functions for different values of 
the interaction radius. The chain length is $N=50$ dipoles and $\mu=30\,$MD units, $\gamma=5.0$. In the left panel, the heat flux is generated by the 
temperature differences of thermostats $T_L=0.1\,$MD units, $T_R=0.3\,$MD units. In the right panel, the space correlation function in equilibrium conditions is presented, $T=0.2\,$MD units.}
\end{figure}

Fig.~\ref{fig:space correalation functions comparison} demonstrates emergence of new long-range correlations in nonequilibrium steady state, which 
are not present in thermal equilibrium. The difference between the equilibrium and nonequilibrium states reaches several orders of magnitude for 
dipoles lying on the distance of five and more lattice sites away from some fixed dipole. The equilibrium correlation function decays fast while the
nonequilibrium one weakly varies on the length scale of the chain.

One more noteworthy peculiarity of space correlations is observed. Altering the range of dipole-dipole interaction (by changing the cutoff radius $N_{cut}$)
strongly affects space correlations in a system. In Fig.~\ref{fig:space correalation functions comparison} we see that the magnitudes of the 
correlations are substantially different for chains with $N_{cut}=2\div4$ and $N_{cut}\geq 7$. In Fig.~\ref{fig:space correalation functions 
comparison} we also show the difference between correlations in equilibrium and nonequilibrium states. It is seen that there is a long range order in 
nonequilibrium for $N_{cut}\neq 2$ that disappears in equilibrium state. Nevertheless, there is some similarity in correlations for a nearest-neighbor 
interaction, but even in this case a presence of the heat flux makes the correlations decay slower. For long chains the effect still persists.

\begin{figure}
\centering
\includegraphics[scale=1.0]{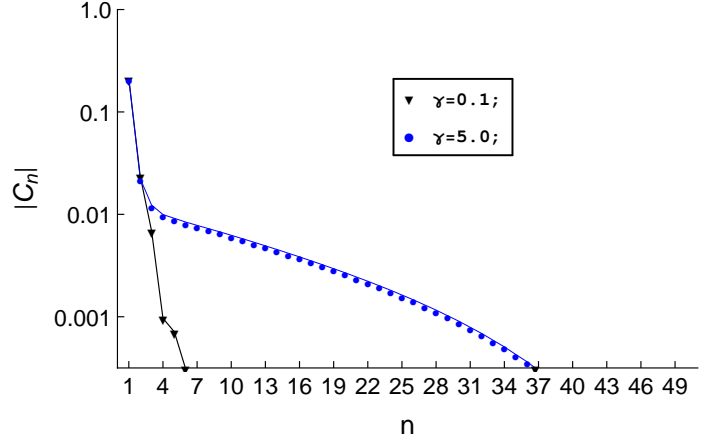}
\caption{\label{fig:different coupling}  The absolute values of space correlation function in strong and weak coupling regimes. The chain length is 
$N=50$ dipoles and $\mu=30\,$MD units, $T_L=0.1\,$MD units, $T_R=0.3\,$MD units, $N_{cut}=9$. }
\end{figure}

This long-range character of the space correlation function is due to the heat flux,and behavior of the equilibrium correlation function 
supports this conjecture. The right graph in Fig.~\ref{fig:space correalation functions comparison} convincingly demonstrates that without heat 
flux, \textit{i.e.}, in an equilibrium state, the space correlation function decays fast with distance for the same set of parameters as for 
nonequilibrium one. The link between the space correlation function and the heat flux originates from the fact that coupling to the thermostats 
results in the interaction between eigenmodes of the chain~\cite{Lepri03}. From Fig.~\ref{fig:different coupling}, one can see the differences between weak 
($\gamma=0.1$) and strong ($\gamma=5.0$) coupling of the chain to thermostats.

\begin{figure}
  \centering
  \begin{subfigure}[b]{0.45\textwidth}
  \includegraphics[width=\textwidth]{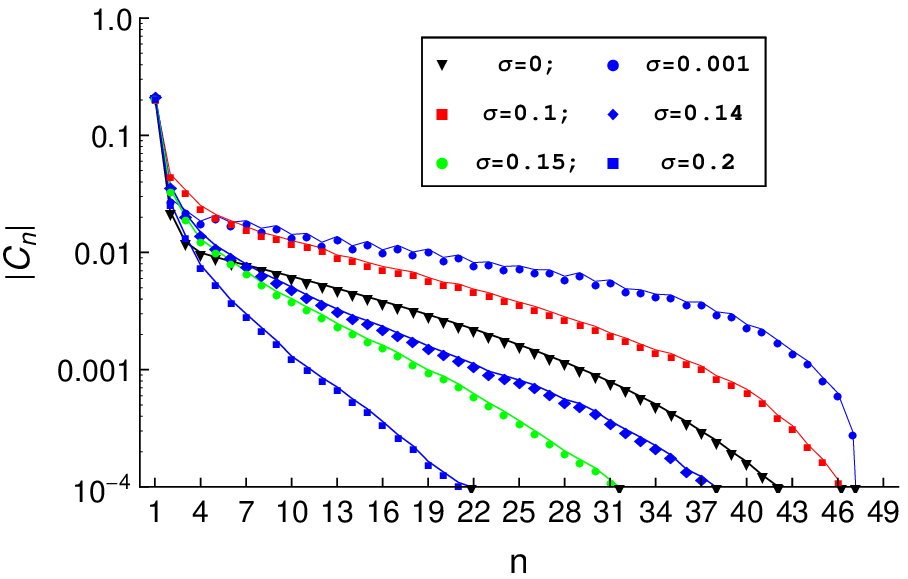}
  \end{subfigure}
  \begin{subfigure}[b]{0.45\textwidth}
  \includegraphics[width=\textwidth]{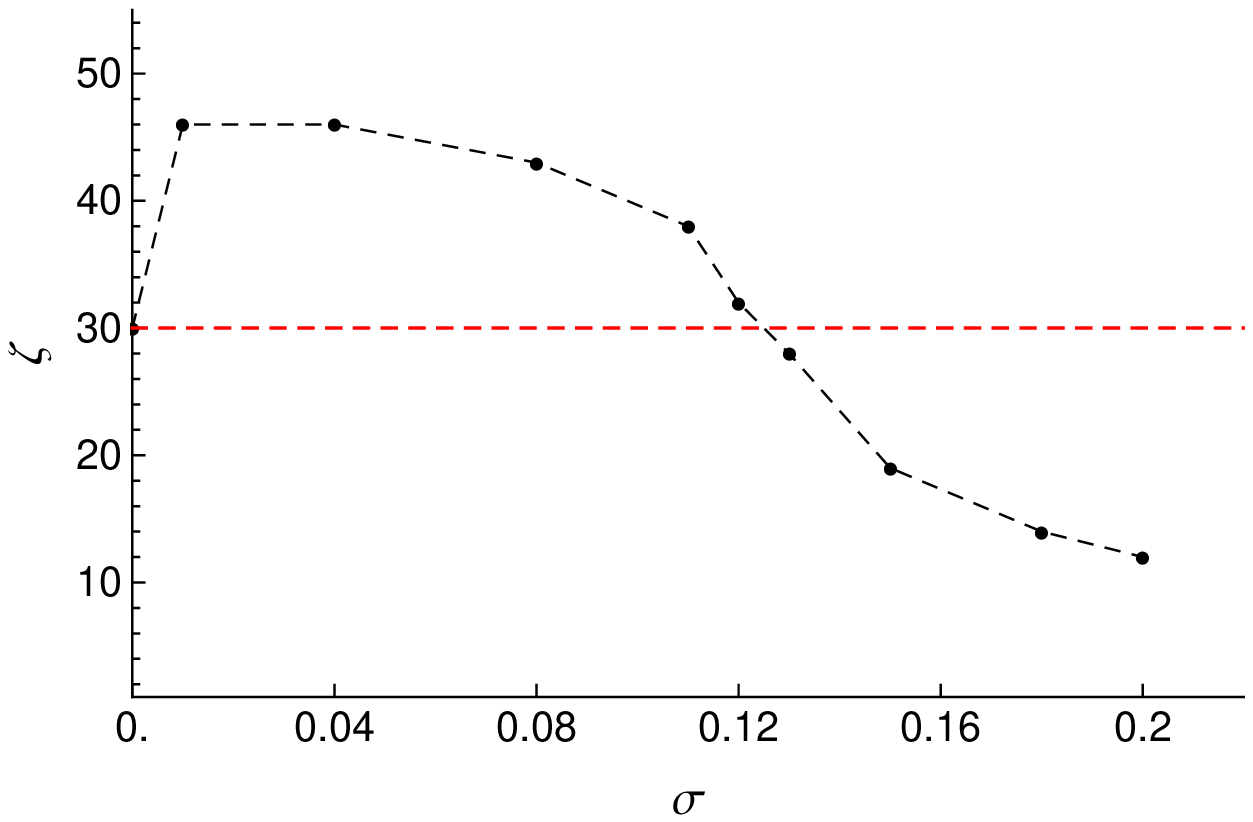}
  \end{subfigure}
\caption{\label{fig:space correlations disorder}  The absolute values of space correlation function for different disorder strengths. The chain length 
is $N=50$ and $\mu=30\,$MD units, $T_L=0.1\,$MD units, $T_R=0.3\,$MD units, $N_{cut}=9$. The right panel shows dependence of the correlation length $\zeta$ on the disorder strength is presented. Horizontal dashed line represents the value of a correlation length in ordered chain, $\sigma=0$.}
\end{figure}

Finally, we consider the influence of the disorder on space correlations. Surprisingly, the effect of the disorder turns out to be not strictly 
''destructive''. From the left panel of Fig.~\ref{fig:space correlations disorder} we see that for $\sigma=0.01,\,0.1$ correlations are stronger than ones in an ordered chain, whereas for higher values of $\sigma$ they become weaker. To quantitatively estimate this result, we introduce correlation length $\zeta$ defined as a minimal number $n_{\zeta}$ such that $|C_{n_{\zeta}}|<10^{-3}$. Nonmonotonic character of $\zeta(\sigma)$ is evidently seen in the right panel of Fig.~\ref{fig:space correlations disorder}. This result is very intriguing because from general considerations disorder should 
break the long-range order in a system. Moreover, it is important to note, that the enhancement of correlations is almost independent on the range of correlations. Correlation functions $C_n$ calculated with $N_{cut}=2$ and $N_{cut}=9$ are close enough to say that effect of disorder weakly depends on $N_{cut}$. Another important point is the absence of this effect in equilibrium conditions. 

\begin{figure}
\includegraphics[scale=1.1]{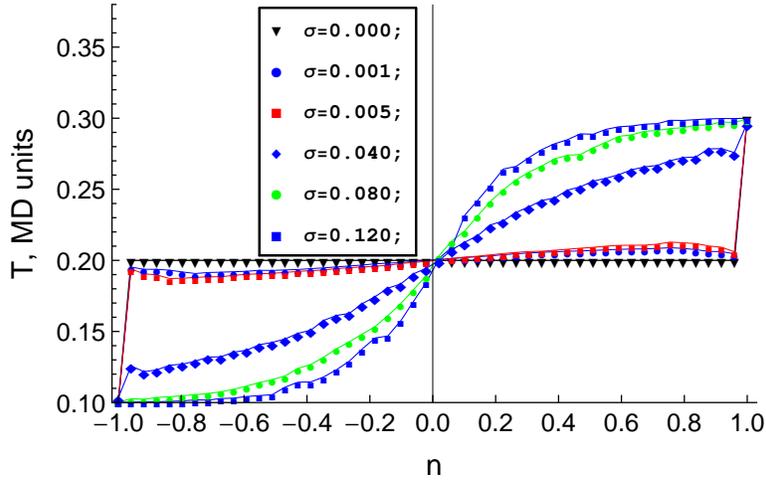}
 \caption{\label{fig:temperature profiles new} Temperature profiles for different disorder strengths according to Eq.~\eqref{eq:temp profile approximation}. Profiles in disordered cases were obtained by averaging over $200$ disorder realizations, $T_1=0.1\,$MD units, $T_2=0.3\,$MD units, $N_{cut}=9$.}
 \end{figure}

To elucidate the origin of this amplification of long-range correlation by the disorder, we computed the local temperature profile along the dipole chain (Fig.~\ref{fig:temperature profiles new}). As one can see from  Fig.~\ref{fig:temperature profiles new}, in ordered case ($\sigma=0$) the temperature is flat along the chain and changes only on the interface dipoles. When we introduce a small disorder ($\sigma<0.12$), a linear temperature gradient develops along the entire chain, which couples all dipoles and leads to the increase of the long range correlation. The further increase of a disorder results in a strong localization of a thermal gradient in the middle of the chain, part of dipoles become ''uncoupled'' and we observe weakening of the correlations. It is known from a nonequilibrium thermodynamics~\cite{deGroot2013} that in polarizable media a temperature gradient creates an electric field that raises overall polarization~\cite{Bresme2008,Muscatello2011}. 
\section{Summary}
In the present paper, we have conducted numerical and analytical study of the classical dipolar chain under out-of-equilibrium conditions. We 
approximated nonequlibrium dynamics of the chain by a system of linearized stochastic differential equations. All the quantities of interest after that were 
expressed solely through the elements of the one matrix $\mathbf{M}$ Eq.~\eqref{eq:M-matrix}. We focused on two basic aspects of the chain: heat 
conduction and correlations.

To study the heat conduction we derived a closed expression for the heat current, Eq.~\eqref{eq:current Langevin}. It was established that ordered dipole 
chain supports ballistic transport and their properties resemble ones of the harmonic lattice; \textit{e.g.}, the heat current is proportional to the 
difference of the thermostats' temperatures $\Delta T$ and not to the temperature gradient $\Delta T/N$. This fact points to the violation of the 
Fourier law in ordered dipolar chain. Ballistic transport regime is destroyed by a disorder introduced by a random distribution of dipoles' positions. 
We used the simple model where positions of dipoles are imposed to be random Gaussian variables. The dispersion of the distribution plays a role of 
the disorder's ''strength''. Within the adopted model of disorder, we calculated the temperature profile and heat conductivity. It was observed that 
heat conduction undergoes the transition from ballistic to diffusive transport. In the diffusive regime, the heat conductivity
decreases as we increase the chain length.
Diffusive regime is also characterized by the establishing of the temperature gradient in chain. Deformation of the temperature profile in disordered 
chain is in good agreement with the recent results on the derivation of the Fourier's law in quantum wires~\cite{Dubi2009}. Similar to the quantum 
case~\cite{Dubi2009}, there are two different length scales in a problem of heat conduction. The first one corresponds to localization length. The 
second one corresponds to the thermal length. These two lengths can be very different. In the model considered in the present article, estimation of 
the localization length is complicated by the presence of the strong correlations between eigenmodes. We left this question for further 
consideration.

We constructed the exact formula for ''dipole-dipole'' correlation function Eq.~\eqref{eq:correlation function analytic}. This allowed us to estimate the relaxation times and to show the slowing of orientation relaxation as the system size increases (see Fig.~\ref{fig:relaxation times}).

The situation with spatial correlations is more subtle because they are affected by different factors such as thermostats' temperature difference, 
dipole moments, and interaction range (cutoff radius $N_{cut}$). The most prominent feature of nonequilibrium is the emergence of the long-range 
correlations for $N_{cut} > 5$. It is especially important because usually the model with only  nearest-neighbor interaction is considered in the majority 
of works in out-of-equilibrium low dimensional systems. Here we clearly demonstrated that long-range behavior of the correlation is caused by the 
combination of two factors: heat flow and ''long-range'' interaction. The nonequilibrium is essential because it leads to the coupling between heat-carrying modes of the systems and the eigenmodes interaction is necessary for the emergence of long-range structure. One can generalize this conclusion by making the conjecture that the emergence of long-range correlations in a one-dimensional system is possible under nonequilibrium conditions when next-nearest-neighbor interactions are included.

\section{Acknowledgments}
This work has been supported by the Belgian Federal Government. The authors thank Maxim Gelin for many valuable discussions.

\appendix

\section{\label{sec:appA} Units of measure adopted in the article}
 \begin{itemize}
 \item The value of the dipole moment corresponds to the one of the water in the carbon nanotube and is taken from the molecular dynamics 
     simulation~\cite{Kofinger2010} : $\mu=1.9975\, D$;
 \item Length unit: $a=2.65$ \AA, \textit{ibid};
  \item Energy unit: $\epsilon=\displaystyle{\frac{\mu^2}{a^3}}$,  with the above values of the dipole moment and lattice spacing we get 
      $\epsilon=2.14406\cdot 10^{-13}\, erg$;
 \item The unit of time: $\tau=\displaystyle{\sqrt{ \frac{I a^3}{\mu^2} }}=3.74161\cdot 10^{-14}\,s$, where $I$ is the moment of inertia of a 
     dipole. For $I$ we take the mean value of the three principal values of this tensor of the water molecule~\cite{Eisenberg2005}, $I =3.0\cdot 
     10^{-40}\, g/cm^2$;
\end{itemize}
\section{\label{sec:appB} Space correlation function in nonequilibrium steady state}
In this section we give some useful formulas that are used for calculation of the space correlation function of one-dimensional lattices~\cite{Rieder1967, Nakazawa70, Casati1979}. Below we closely follow the work of Ref.~\cite{Nakazawa70}.

We start with rewriting the system Eq.~\eqref{eq:linearized Langevin dynamics} in the form of the first order differential equation:
\begin{equation}\label{eq:first order dynamics}
\dot{\mathbf{x}} = \mathbf{A} \mathbf{x} + \mathbf{F}(t),
\end{equation}
where the column vector $\mathbf{x}(t) = (\varphi_1(t),\ldots,\varphi_N(t), p_1(t)\ldots p_N(t))^T$ and
\begin{equation}\label{eq:big matrices}
\begin{aligned}
A=\begin{pmatrix}
  \mathbf{0}_{NN} & \phantom{-}\mathbf{I}_{NN} \\
  \mathbf{M}               & -\mathbf{\Gamma}
  \end{pmatrix}, \quad
  \Gamma_{ij}=\gamma_L \delta_{i1}\delta_{j1} + \gamma_R \delta_{iN}\delta_{jN}
  \end{aligned}
\end{equation}
where $\mathbf{0}_{NN}$ is $N$-by-$N$ zero matrix, $\mathbf{I}_{NN}$ is $N$-by-$N$ identity matrix, $\mathbf{M}$ is the potential energy matrix, 
$\mathbf{F}=(\mathbf{0}_N,\eta_L(t),0,\ldots, \eta_R(t) )^T$ and $\eta_{L}(t)$, $\eta_{R}(t)$ are Gaussian white noises. Solution of Eq.~\eqref{eq:first order dynamics} is of the form 
\begin{equation}\label{eq:solution of the first order dynamics}
\mathbf{x}(t)=\exp(\mathbf{A}t)\mathbf{x}(0) + \int\limits_{0}^{t}\exp\Big(\mathbf{A}(t-t')\Big)\mathbf{F}(t')\,dt',
\end{equation}
where $\mathbf{x}(0)$ is column of the initial conditions. To find the correlations in the nonequlibrium steady state we have to calculate the limit 
$\lim_{t\to+\infty}\langle \mathbf{x}(t)\mathbf{x}^{\dagger}(t)\rangle$, where $\langle \mathbf{x}(t)\mathbf{x}^{\dagger}(t)\rangle$ is the covariance 
matrix. It can be done by employing the fact that all eigenvalues of the matrix $\mathbf{A}$ have negative real parts~\cite{Rieder1967,Nakazawa70} 
which gives $\lim_{t\to+\infty} \exp(\mathbf{A}t)=0$. After this remark the evaluation of the limit above is done in a straightforward manner:
\begin{equation}\label{eq:long time limit of correlation function}
\lim_{t\to+\infty}\langle \mathbf{x}(t)\mathbf{x}^{\dagger}(t)\rangle=
\begin{pmatrix}
\boldsymbol{\Phi} & \mathbf{Z} \\
\mathbf{Z}^{\dagger} & \mathbf{T}
\end{pmatrix}
=\int\limits_0^{\infty}\exp(\mathbf{A}t)\mathbf{D}\exp(\mathbf{A}^{\dagger}t)\,dt,
\end{equation}
where
\begin{equation}
\begin{aligned}
&\langle \mathbf{F}(t)\mathbf{F}^{\dagger}(t')\rangle =\mathbf{D}\delta(t-t')=
\begin{pmatrix}
\mathbf{0}_{NN} & \mathbf{0}_{NN} \\
\mathbf{0}_{NN} & \mathbf{\Delta}
\end{pmatrix}\delta(t-t'), \quad \Delta_{ij} = 2\gamma_L T_L \delta_{i1}\delta_{j1} + 2\gamma_R T_R \delta_{iN}\delta_{jN}, \\
&\boldsymbol{\Phi} = \langle \boldsymbol{\varphi} \boldsymbol{\varphi}^{\dagger}\rangle,\quad \mathbf{Z} = \langle \boldsymbol{\varphi} 
\boldsymbol{p}^{\dagger}\rangle, \quad \mathbf{T} = \langle \boldsymbol{p} \boldsymbol{p}^{\dagger}\rangle.
\end{aligned}
\end{equation}
Here we have used the fluctuation-dissipation theorem Eq.~\eqref{eq:fluctuation dissipation} and definition of $\mathbf{F}(t)$ given above. The matrices 
$\boldsymbol{\Phi}$, $\mathbf{Z}$ and $\mathbf{T}$ represent space correlation function, mean heat flux, and temperature profile (diagonal elements of 
$\mathbf{T}$) respectively. The correlation function can be obtained by direct calculation of the integral Eq.~\eqref{eq:long time limit of correlation 
function} by the method developed in Ref.~\cite{vanLoan1978}. It can be also helpful to rewrite Eq.~\eqref{eq:long time limit of correlation function} in the matrix 
form~\cite{Rieder1967,Nakazawa70}:
\begin{equation}
\mathbf{A}\mathbf{C} +  \mathbf{C}\mathbf{A}^{\dagger}=-\mathbf{D},
\end{equation}
with $\mathbf{C}$ being $\langle \mathbf{x}(+\infty)\mathbf{x}^{\dagger}(+\infty)\rangle$. An exact solution of this equation was found only for the 
nearest-neighbor interaction. From a general point of view, it is a well known in a control theory Lyapunov matrix equation and number of algorithms 
were developed to solve it numerically~\cite{Golub1979, Wachspress1988,Penzl1998}.

\section{\label{sec:appC} The inverse of the matrix \textbf{M} and calculation of the heat current}
In the Langevin dynamics we calculate two main quantities: heat current, Eq.~\eqref{eq:current Langevin}, and the correlation 
function, Eq.~\eqref{eq:correlation function Langevin}. In both cases, the formulas contain the elements of the inverse matrix $\mathbf{M}^{-1}$ that, by 
definition, is given by
\begin{equation}\label{eq:general inverse}
\mathbf{M}^{-1}=\frac{1}{\det \mathbf{M}} \mathbf{C}^{\mathrm{T}},
\end{equation}
where $\mathbf{C}$ is the matrix of cofactors~\cite{Gantmacher1960} and $\mathrm{T}$ stands for transposition. Due to the $\mathbf{M}$ being symmetric a
matrix, the matrix of cofactors $\mathbf{C}$ is also symmetric.

The integrands in Eq.~\eqref{eq:correlation function Langevin} and Eq.~\eqref{eq:current Langevin} are the rational functions and the order of the polynomial in 
the denominator is $2N$ (see comments at the end of the Sec.~\ref{subsec:Langevin dynamics}). From general considerations, it follows that there 
should be no zeros of the $\det{\mathbf{M}}(\omega)$ on the real axis because if there were even one the integral in Eq.~\eqref{eq:current Langevin} 
became ambiguous while the heat current is a physically observable quantity and must be defined unambiguously.

\begin{figure}
\includegraphics[scale=0.9]{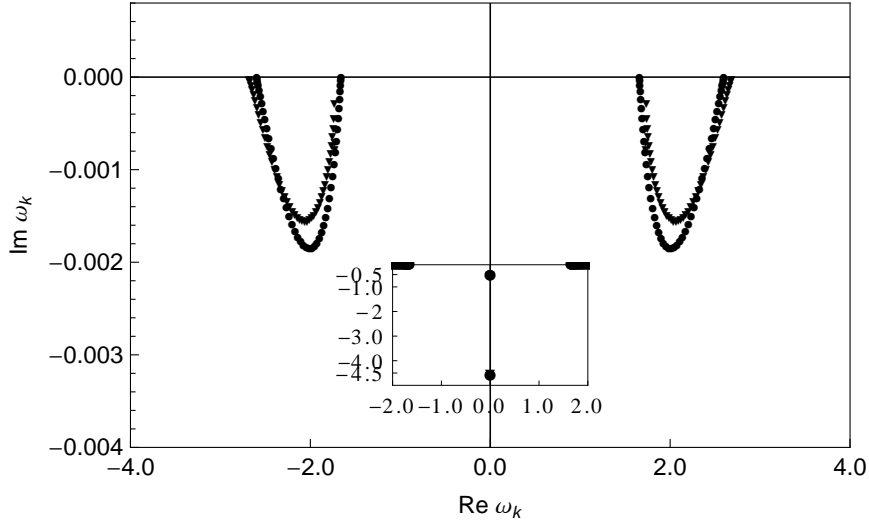}
 \caption{\label{fig:polesVsNc} Zeros of $\det \mathbf{M}(\omega)$ for different $N_{cut}$; $\bullet$ correspond to $N_{cut}=3$, $\blacktriangledown$ 
 correspond to $N_{cut}=20$; $N=50$, $\mu=1.0\,$MD units}
 \end{figure}

Now, we are ready to calculate the integral
\begin{equation}
\label{eq:current Langevin residues}
\langle j \rangle = \frac{2\gamma_L \gamma_R \Delta T }{2\pi} \int\limits_{-\infty}^{+\infty}\omega^2 |(M^{-1})_{1N}(\omega)|^2\,d\omega.
\end{equation}
From the residue theorem, it immediately follows~\cite{Ablowitz2003}:
\begin{equation}\label{eq:current residue theorem}
\langle j \rangle=2\pi i\frac{2\gamma_L \gamma_R \Delta T }{2\pi}\sum\limits_{\omega_i}\mathop{\mathrm{Res}}\limits_{\omega=\omega_i}\left( 
\omega^2\frac{\mathbf{C}_{N1}(\omega)\mathbf{C}_{N1}(-\omega)}{\det\mathbf{M}(\omega)\det\mathbf{M}(-\omega)}\right),
\end{equation}
where $\omega_i$ satisfies the equation
\begin{equation}\label{eq:zeros of the determinant}
\det\mathbf{M}(\omega_i)=0,
\end{equation}
in the half-plane $\mathrm{Im}\,\omega<0$ and $\mathop{\mathrm{Res}}\limits_{\omega=\omega_i}$ states for the residue at the point $\omega=\omega_i$. 
The determinant can be represented in the form of a product:
\begin{equation}\label{eq:determinant form}
 \det\mathbf{M}(\omega)=\prod\limits_{i=1}^{2N}(\omega-\omega_i).
\end{equation}
It is tacitly supposed that roots of this polynomial are simple. This assumption is based on the results of the numerical computations: for all 
considered chain lengths, the roots are found to be simple and all lie in the lower half-plane of the $\omega$-plane, Fig.~\ref{fig:polesVsNc}. Hence,
poles in Eq.~\eqref{eq:current residue theorem} are simple and calculation of the sum can be done easily. To be assured we checked the teh multiplicity of 
the roots in the Multroot package~\cite{Zeng2004} yielded the same conclusion.

At the end of the section we will show how to tackle the root-finding of the $\det \mathbf{M} (\omega)$. As it was already said, this polynomial has 
large coefficients and is of the order of $2N$. Hence, it is a cumbersome problem to find its roots. Nevertheless, it can be greatly simplified by 
the following observation.

Applying the Fourier transform to Eq.~\eqref{eq:first order dynamics}, we get the matrix equation:
\begin{equation}
-\ci \omega\mathbf{x}(\omega)=\mathbf{A} \mathbf{x}(\omega) + \mathbf{F}(\omega).
\end{equation}
Thus, a particular solution of Eq.~\eqref{eq:first order dynamics} is
\begin{equation}
\mathbf{x}(t)=\frac{1}{2\pi}\int\limits_{-\infty}^{+\infty}\exp(-\ci\omega t)\mathbf{x}(\omega)\, d\omega = -\frac{1}{2\pi}\int\limits_{-\infty}^{+\infty}\frac{\exp(-\ci\omega 
t)}{\ci\omega\mathbf{I}+\mathbf{A}}\mathbf{F}(\omega)\,d\omega,
\end{equation}
where $\mathbf{I}$ is $2N$-by-$2N$ identity matrix. Comparison of this expression with the inverse Fourier transform Eq.~\eqref{eq:matrix equation for angles} shows that  zeros of $\det\mathbf{M}(\omega)$ coincide 
with zeros of $\det (\ci\omega \mathbf{I} + \mathbf{A})$ and the latter are equal to $\ci\lambda_k$, $k=1,2\ldots 2N$, where $\lambda_k$ is the $k$-th 
eigenvalue of the $\mathbf{A}$. Therefore,
\begin{equation}\label{roots of polynomial vs eigenvalues}
\det \mathbf{M}(\omega_k) = 0, \quad \forall\,\omega_k = \ci\lambda_k, \; k=1,2,\ldots,2N.
\end{equation}
According to this observation, the determinant of the $\mathbf{M}(\omega)$ in Eq.~\eqref{eq:current residue theorem} can be evaluated as
\begin{equation}\label{eq:determinant vs eigenvalues}
\det\mathbf{M}(\omega)=\prod\limits_{k=1}^{2N}(\omega-\mathrm{i}\lambda_k).
\end{equation}
The problem of finding eigenvalues of the nonsingular matrix is well known and can be implemented in a robust and reliable way.

Now, when we know that all zeros of $\det \mathbf{M}(\omega)$ are simple and know the relation between them and eigenvalues of $\mathbf{A}$, we can 
calculate the residue in Eq.~\eqref{eq:current residue theorem}:
\begin{equation}\label{eq:current residue theorem explicit}
\langle j \rangle=2\ci\gamma_L \gamma_R \Delta T \sum\limits_{n=1}^{2N} 
(\ci\lambda_n)^2\frac{\mathbf{C}_{N1}(\ci\lambda_n)\mathbf{C}_{N1}(-\ci\lambda_n)}{\det'\mathbf{M}(\ci\lambda_n)\det\mathbf{M}(-\ci\lambda_n)},
\end{equation}
where $\det\nolimits'\mathbf{M}(\ci \lambda_n)=\ci\prod\limits_{\substack{k=1 \\ k\neq n}}^{2N}(\lambda_n-\lambda_k)$.

\section{Energy current in dipole chain}
We begin with  the rate of the energy change in chain~\cite{Lepri03, Dhar08}:
\begin{equation}
\frac{dE}{dt} = \sum\limits_{i=1}^N \dot{p_i}p_i + \frac12\sum\limits_{i,j}^N\left( \frac{\partial U}{\partial \varphi_i}\dot{\varphi_i}  + 
\frac{\partial U}{\partial \varphi_j}\dot{\varphi_j} \right) = \sum\limits_{i,j}F_{ij}p_i - \frac12\sum\limits_{i,j}^N\left( F_{ij}p_i + F_{ji}p_j  
\right).
\end{equation}
The energy change rate for the $i$-th dipole is:
\begin{equation}
 \frac{d\epsilon_i}{dt}=\frac12 p_i\sum\limits_j F_{ij} - \frac12\sum\limits_j F_{ji}p_j.
\end{equation}
Rewrite it in the following form:
\begin{equation}
\frac{d\epsilon_i}{dt} + \left(j_{i}^{in} - j_{i}^{out}\right)=0,
\end{equation}
where
\begin{equation}
j_i^{in}   = \frac12 p_i \sum\limits_{j=1}^N F_{ij},\quad
j_i^{out} = \frac12 \sum\limits_{j=1}^N F_{ji}p_j.
\end{equation}

In steady state, the $\left\langle \frac{d\epsilon_i}{dt} \right\rangle=0$, thus $\langle j_i^{in} \rangle = \langle j_i^{out} \rangle$ for every 
index $i$, where $\langle \phantom{x} \rangle$ stands for the ensemble average. The  $ j_i^{in}$, $ j_i^{out}$ can be treated as the \textit{in, out} 
- heat currents of $i$th dipole, respectively.  In a straightforward manner it can be shown~\cite{Lepri03, Dhar08} that ''in'' steady state \textit{in}-currents are 
equal for all dipoles in chain:
\begin{equation}\label{eq:energy current}
j_2=\frac12 p_2 F_{21}=j_3=\frac12 p_3 \sum\limits_{i=1}^2 F_{3i}=\ldots=j_N=\frac12 p_{N-1}\sum\limits_{j=1}^{N-1}F_{N-1,j}.
\end{equation}
Evidently, all these currents in steady state are equal to the energy flowed into the system from the ''hot'' thermostat per unit of time and flowed 
out into the ''cold'' thermostat.
Equation~\eqref{eq:energy current} states that  the average rate of work done by the $i$th dipole on the followed dipoles  is equal to the rate of 
work done by the previous dipoles on the $i$th. Thus, one can say that the energy flow from the previous dipoles to the $i$th is equal to the energy 
flow from the $i$th dipole to the followed ones.

\end{document}